%% file: 99-main.tex
\documentclass{article}
\input{99-boilerplate}

\isAnonymousfalse

\input{00-title}

\ifisAnonymous
\else
  \input{00-authors}
\fi  %

\begin{document}

\maketitle

\input{00-abstract}
\input{01-introduction}
\input{02-example}
\input{03-vision}

\input{04-related}
\input{05-broader-perspective}
\ifisAnonymous
\else
  \input{07-acks} 
\fi  %

\printbibliography

\end{document}

%% file: 99-boilerplate.tex
\usepackage[arxiv, nonatbib, final]{neurips_2026}

\usepackage{array}
\usepackage{ragged2e}
\usepackage{tikz}
\usepackage{amsmath}
\usepackage{hyperref}
\usepackage[normalem]{ulem}
\usepackage{listings}
\usepackage{xspace}
\usepackage{booktabs}
\usepackage{multirow}
\usepackage{wasysym}
\usepackage{caption}
\usepackage{subcaption}
\usepackage{enumitem}
\usepackage[utf8]{inputenc}
\usepackage[compact, small]{titlesec}
\usepackage{tcolorbox}
\usepackage{wrapfig}
\tcbuselibrary{theorems}

\newif\ifisAnonymous

\newtcbtheorem[number within=section]{mytheo}{\small API Call}%
{colback=green!5,colframe=green!35!black,left=1mm,top=1mm,bottom=1mm}{th}

\usepackage[
  backend=biber,
  style=numeric-comp,
  minalphanames=3,
  isbn=false,
  sortcites=true,
  sorting=anyt,
  abbreviate=false,
  url=false,
  doi=false,
  maxnames=99,
  minbibnames=3,
  maxbibnames=99]{biblatex}
\AtBeginBibliography{\small}
\newcommand{\commentswitch}[1]{\xspace}
\renewcommand{\commentswitch}[1]{#1}

\newcommand{\eg}{{e.g.},\xspace}
\newcommand{\ie}{{i.e.},\xspace}

\definecolor{codegreen}{rgb}{0,0.4,0}
\definecolor{codegray}{rgb}{0.5,0.5,0.5}
\definecolor{codepurple}{rgb}{0.58,0,0.82}
\definecolor{backcolour}{rgb}{0.95,0.95,0.92}

\lstdefinestyle{rust}{
    commentstyle=\color{codegreen},
    keywordstyle=\color{codepurple},
    stringstyle=\color{blue},
    basicstyle=\ttfamily\scriptsize,
    breakatwhitespace=false,
    breaklines=true,
    captionpos=b,
    keepspaces=true,
    showspaces=false,
    showstringspaces=false,
    showtabs=false,
    tabsize=2
}

%% file: 00-title.tex
\title{Engineering Robustness into Personal Agents with the AI Workflow Store}

%% file: 00-authors.tex
\author{%
  Roxana Geambasu\thanks{Work done exclusively in the context of Google engagement. Authors ordered alphabetically.} \\
  Columbia University and Google\\
  \texttt{roxanageambasu@google.com} \\
  \And
  Mariana Raykova\\
  Google\\
  \texttt{marianar@google.com}\\
  \And
  Pierre Tholoniat\\
  Google\\
  \texttt{ptholoniat@google.com}\\
  \And
  Trishita Tiwari\\
  Google\\
  \texttt{trishitatiwari@google.com}\\
  \And
  Lillian Tsai\\
  Google\\
  \texttt{tslilyai@google.com}\\
  \And
  Wen Zhang\\
  Google\\
  \texttt{wnz@google.com}\\
}  %

%% file: 00-abstract.tex
\begin{abstract}

The dominant paradigm for AI agents is an ``on-the-fly'' loop in which  agents synthesize plans and execute actions within seconds or minutes in response to user prompts. We argue that this paradigm short-circuits disciplined software engineering (SE) processes---iterative design, rigorous testing, adversarial evaluation, staged deployment, and more---that have delivered the (relatively) reliable and secure systems we use today. By focusing on rapid, real-time synthesis, are AI agents effectively delivering users improvised prototypes rather than systems fit for high-stakes scenarios in which users may unwittingly apply them?
This paper argues for the need to integrate rigorous SE processes into the agentic loop to produce production-grade, hardened, and
deterministically-constrained agent \emph{workflows} that substantially
outperform the potentially brittle and vulnerable results of on-the-fly synthesis.
Doing so may require extra compute and time, and if so, we must amortize the cost
of rigor through reuse across a broad user community.
We envision an {\em AI Workflow Store} that consists of hardened and reusable
workflows that agents can invoke with far greater reliability and security than
improvised tool chains. We outline the research challenges of this vision, which
stem from a broader flexibility-robustness tension that we argue requires moving
beyond the ``on-the-fly'' paradigm to navigate effectively.

\end{abstract}

%% file: 01-introduction.tex
\section{Introduction}
\label{sec:introduction}

AI agents today can exhibit striking failures, \eg
deleting an entire inbox when asked to remove a confidential
message~\cite{rogue3}; erasing a codebase to ``fix'' an authorization
issue~\cite{rogue1}; and compromising developers' machines because of a single
GitHub title containing a prompt injection~\cite{rogue5}. This recalls the web
and software landscape of two decades ago, when frequent crashes and SQL
injection attacks made reliability and security constant concerns. Over time,
disciplined software engineering (SE) and robust frameworks tamed that
landscape, producing systems that (despite caveats) provide enough reliability
and security that we now often take their robustness for granted.

\begin{figure}[h]
    \centering
    \includegraphics[width=.8\linewidth]{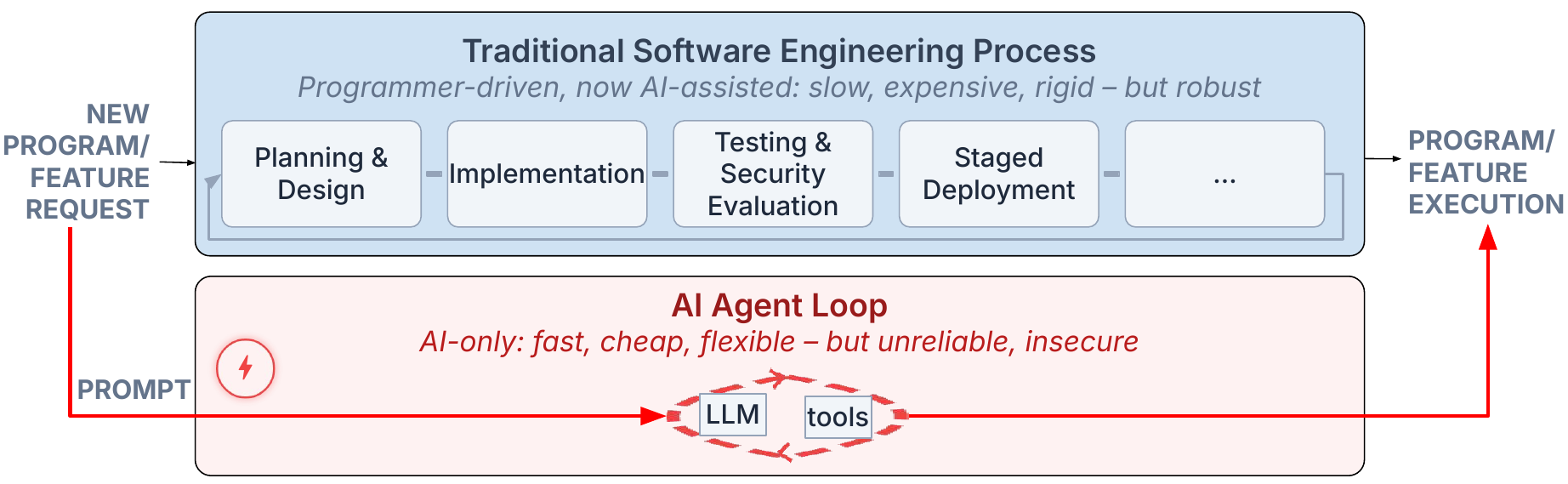}
    \caption{\small Problem: The Agentic AI code-and-execute loop short-circuits well-trodden SE processes that are the foundations of the relatively reliable and secure programs and services we enjoy today.}
    \label{fig:ai-swe-shortcircuit}
\end{figure}

Agentic AI signals a major and largely unvetted transition
(Figure~\ref{fig:ai-swe-shortcircuit}): from highly engineered software toward
agents that synthesize and execute arbitrary functionality on the fly in
response to individual user prompts---a common pattern for {\em personal AI
agents}~\cite{chatgpt_agent, gemini_agent, openclaw_ai}. While these agents can
reason, orchestrate tools, and complete tasks beyond conventional software,
their successes coexist with well-documented failures: brittleness to prompt
changes, incorrect tool invocation, hallucinations, and vulnerability to attacks
such as context poisoning and prompt injection~\cite{willisonpi, rogue1, rogue2,
rogue3, rogue4, rogue5}. 

These failures arise from the substantial ask of the ``on-the-fly'' loop: in
seconds or minutes, and often for pennies, it must synthesize and execute
multi-step plans: sending emails, moving money, booking travel, editing
documents, and coordinating across services in ways that directly affect user
data, accounts, and relationships. In the traditional world, such integrations
would undergo weeks of processes such as design, implementation, testing and
security evaluation, internal beta, and staged rollout before reaching users.
Anything produced ``instantly'' and without these safeguarding processes would
have been labeled a makeshift prototype, and not pushed into production. While
traditional software is far from flawless, its failure and exploitation rates
set the minimum bar for what we can consider ``reliable'' and ``secure.'' %

We argue that in order to reach this bar, on-the-fly agent executions
should be hardened and deterministically constrained into \emph{workflows}
by incorporating systematic SE processes into the agentic loop, including:
requirement collection, design, artifact implementation, testing, adversarial
evaluation, and staged deployment---all supported by robust frameworks for
fuzzing, rollouts, monitoring, feedback, and others. 
To believe that an AI model---however capable---can reliably and securely
synthesize and execute complex plans under acute time and resource
constraints is to reject a central lesson of forty years of software
engineering: \emph{robustness is an engineered property achieved through
rigorous process}, not bestowed by any single component or mind. 

Recent work has begun to reintroduce fragments of SE practice---improved
planning~\cite{chen2025secalign, instructionhierarchy,zverev2025aside}, handling ambiguity~\cite{camber}, security
checks~\cite{conseca, fides, camel, ace, ro2025sherlock}---but these remain sharply constrained by latency and compute cost.
Even seconds of extra reasoning per step are often treated as
prohibitive in a system optimized for immediate response~\cite{latency_ms,
latency_tealium, latency_google}.
However, we hypothesize that achieving workflows that meet the reliability and
security bar may require \emph{far more time and resources} than an on-the-fly
execution---a major burden that incentivized these on-the-fly agent loops to
begin with.
We posit that these SE overheads can be (1) made orders of magnitude faster by
AI automation compared to human-driven development; and (2) made tolerable via \emph{reuse}: individual prompts and
contexts may be unique, but the workflows that satisfy them may be reusable
across users and time.
We hypothesize that we can sufficiently generalize and parameterize the same
underlying workflow to satisfy user tasks that differ only in their
particulars---while still maintaining strong security and reliability
guarantees. 
This mirrors a long-standing lesson from computer systems: caching is effective because
requests are far less unique than they appear. If similar reuse opportunities exist for agent workflows, then the price of proper engineering need
not be paid per prompt, either in latency or computing cost.
For example, ``book the Airbnb Bob recommended,'' ``move \$500 to my savings
account,'' and ``check for meetings from my emails'' represent tasks that many
users will likely wish to perform, but for which users require specific variants
(\eg which booking details, how much to transact from which accounts,
which meetings exist). While expanding a workflow to support variants
relaxes some constraints (\eg the workflow might allow booking either at 
Hyatt or an AirBnb), hardening by SE processes still deterministically
prevents worst-case outcomes to an acceptable level (\eg execution of a
generalized workflow for ``booking hotels'' cannot result in deletion of files).

\textbf{Our vision, which we term the {\em AI Workflow Store}, is to shift AI agents from ad hoc, on-the-fly behavior to operating over {\em workflows}---hardened, deterministically constrained artifacts produced and continuously maintained by automated software engineering processes. Rather than improvising new programs per prompt, agents select and execute vetted workflows that encode safe, reliable patterns of behavior. These workflows accumulate in a shared repository, transforming the one-time cost of rigorous engineering into a durable asset whose cost amortizes across many users and requests. This vision fundamentally matures the modern agentic concept of ``skills''~\cite{AgentSkills} by establishing a concrete ecosystem in which these capabilities are robustly and automatically engineered, discovered, and reused as the norm for agent execution.}

The remainder of this paper develops this vision. \S\ref{sec:example} provides a concrete example---the prompt ``book the Airbnb Bob recommended''---comparing the current on-the-fly paradigm with our proposed AI Workflow Store. \S\ref{sec:vision} envisions the AI Workflow Store's architecture and the research challenges of its design, and \S\ref{sec:related} overviews related efforts and visions.
Finally, \S\ref{sec:broader-perspective} provides a broader perspective on the emerging flexibility-robustness tension in agentic AI and argues that navigating it requires moving beyond on-the-fly improvisation toward engineered workflows.

%% file: 02-example.tex
\section{A Motivating Example}
\label{sec:example}

Consider a personal AI agent handling the prompt: \emph{``Book the Airbnb hotel that Bob recommended in his recent email.''} It is simple yet
realistic; consequential enough to matter (the agent spends the user's money and decides where the user sleeps); and rich enough to expose the gap between on-the-fly execution and engineered workflows. We discuss three settings: a vanilla on-the-fly agent and its vulnerability to prompt injection (\S\ref{sec:example:vanilla}); an on-the-fly agent with a representative in-loop defense and that defense's limits (\S\ref{sec:example:defense}); and the kind of workflow the AI Workflow Store might produce, which sidesteps the vulnerability altogether (\S\ref{sec:example:engineered}). Our point is not that the last is the ``right'' design, but that a properly resourced, off-the-fly path can surface and evaluate options invisible on the fly---some of which may turn the table on adversaries.

\subsection{Vanilla on-the-fly execution}
\label{sec:example:vanilla}

A vanilla agent has access to the user's inbox, a booking tool, and many other tools needed for unrelated prompts. It searches the inbox for Bob's recommendation, extracts the relevant details (location, dates, listing URL), and invokes the booking tool. This works well in most cases. But two failure modes are immediate and well documented. First, even without an adversary, the agent may pick the wrong email (multiple Bobs or recommendations), extract the wrong parameters (confusing dates across a thread), or delete all emails~\cite{openclaw-delete-emails}. Second, and more seriously, the inbox is an \emph{untrusted channel}: any email can carry adversarial content. A malicious message encountered during search can redirect the agent to forward the inbox to an attacker, book a different property, exfiltrate personal data under the guise of ``clarifying'' the request, or reach for any other tool at its disposal---transferring funds, deleting files, and so on. Such \emph{prompt injection} attacks are well documented~\cite{nassi2025invitationneedpromptwareattacks,ignore,indirectpi}.

\subsection{An on-the-fly defense, and why it is not enough}
\label{sec:example:defense}

Suppose the agent execution runs with on-the-fly constraints generated upon every user prompt (\eg Conseca~\cite{conseca} policies or CaMeL~\cite{camel} one-shot code).
These solutions restrict tool accesses based on trusted signals such as the user's prompt and data provenance. This mitigates part of the problem: the agent can no longer be prompt-injected into invoking clearly out-of-scope tools, such as file-system or cross-site tools used for exfiltration or deletion. But other limitations and vulnerabilities can remain.
First, both these solutions generate constraints on agent execution up-front
from the initial context before the workflow unfolds. The lack of full context
requires the policy or script generator to make assumptions can under- or
over-restrict behavior: if Bob's email links to a Drive document containing the
real booking details, these
approaches cannot retroactively grant Drive access.
Furthermore, the simplicity of what constraints can be generated on-the-fly
result in lack of hardened checks: a quick query to an LLM to generate a script
for this user task (as done in CaMeL's approach,
Listing~\ref{lst:camel-workflow}) results in an un-engineered (and untested)
data extraction prompt and the brittle assumption that the queried email refers
to the correct (rather than a malicious) Bob.

\input{02.1-example-on-the-fly-defense}

\subsection{What an engineered workflow might look like}
\label{sec:example:engineered}

Suppose a disciplined SE process is applied to the same prompt. A team
performing requirements gathering, threat modeling, design exploration, and
adversarial testing may reach designs an on-the-fly process is unlikely to
consider. Such designs might offer significant security benefits and should be
evaluated in production to determine the optimal tradeoff between security and
usability according to a target metric.

For example, a different design with a much reduced attack
surface would incorporate a visual overlay directly into the email client, akin
to Gmail's ``Add to Calendar'' feature. As shown in
Listing~\ref{lst:engineered-workflow}, an LLM classifies incoming emails; when
it recognizes an accommodation recommendation from a carefully curated set of
trusted providers, a ``Book this Recommendation'' link appears. In this model,
the user chooses exactly which booking to make; such user involvement~\cite{DBLP:journals/corr/abs-2502-02649} ensures the agent does not have
to guess or search through an inbox of potentially malicious emails that could
trigger a prompt injection causing inbox exfiltration or other dramatic events.
A hijacked classification at worst now surfaces a misplaced link, not a
misdirected wire transfer. Other risks may remain---the LLM may still extract
incorrect parameters, for example---but disciplined SE processes can test and
refine model behavior and prompts \emph{specific to this workflow}, improving it
as production experience accumulates across users. While models are continuously
improved in general through retraining and reinforcement learning, ironing out
bugs for specific use cases can more directly improve reliability and security
for those cases.

We do {\em not} claim the above is the ``right'' design: it changes the interaction
pattern and it is unclear how it may impact usability if more and more overlays
like this accumulate. Our point is that we cannot know---and on-the-fly agents
also cannot know---until the design is implemented and evaluated, alongside other
designs, in production. Our vision is that having the possibility of performing
such design explorations and evaluations is important for providing users with
workflows that provide the best security/usability tradeoffs---and it is what
the on-the-fly loop alone excludes.

%% file: 02.1-example-on-the-fly-defense.tex
\begin{figure*}[t]
\begin{minipage}[b]{0.45\textwidth}
\begin{lstlisting}[
  language=Python, 
  frame=none, 
  basicstyle=\footnotesize\ttfamily,
  columns=flexible, 
  breaklines=true, 
  numbers=none,
  caption={\small On-the-fly, LLM-generated script given the user request (assumes tools exist).},
  label={lst:camel-workflow}
]
def book_airbnb():
  email = gmail_tool.search_messages(
    query="from:Bob airbnb")
  if not email:
    return "No relevant email from Bob"
  
  prompt = f"""Extract the email's Airbnb booking details. Return ONLY a JSON object with: listing_id, check_in_date, and check_out_date. Email: {email}"""
  data = llm.generate_json(prompt) 
  
  try:
    confirmation = airbnb_tool.book(
      data["listing_id"], 
      data["check_in_date"], 
      data["check_out_date"])
    return confirmation
  except Exception as e:
    return f"Booking failed: {str(e)}"
\end{lstlisting}
\end{minipage}
\hfill
\vrule depth -1ex 
\hfill
\begin{minipage}[b]{0.51\textwidth}
\begin{lstlisting}[
  language=Python, 
  frame=none, 
  basicstyle=\footnotesize\ttfamily,
  columns=flexible, 
  breaklines=true, 
  numbers=none,
  caption={\small Pseudocode for an engineered workflow supporting booking only through trusted sites.},
  label={lst:engineered-workflow}
]
# Runs on receipt of each incoming email.
def onReceiptOfEmail(email):
  # LLM classifies the email as an 
  # accommodation recommendation.
  is_recommendation = llm.prompt(
    email, ACCOMMODATION_RECOMMENDATION_PROMPT)
  if not is_recommendation:
    return

  # LLM extracts booking parameters 
  # from the recommendation email.
  params = llm.prompt(
    email, EXTRACT_PARAMS_PROMPT(
        ["location", "dates", "url"]))

  # Only surface the overlay for trusted, 
  # supported booking sites.
  if is_trusted_booking_site(params["url"]):
    ui.render_booking_overlay(email, params)

# Runs only if the user clicks the 
# "Book this Recommendation" button.
def onBookingOverlayClick(params):
  booking_api = get_booking_api_from_url(
    params["url"])
  booking_api.book(params)
\end{lstlisting}
\end{minipage}
\vspace{-10pt}
\end{figure*}

%% file: 03-vision.tex
\section{Vision: AI Workflow Store}
\label{sec:vision}

The engineered design above is what the AI Workflow Store is meant to produce.
Our vision hinges on two {\bf hypotheses}: (1) that AI-driven software
engineering processes, properly resourced, can produce hardened and constrained
workflows that will lead to more reliable and secure agent operation compared to purely on-the-fly operation; 
and (2) that across a population of users, the possible cost of engineering workflows can be amortized through reuse.
After sketching our architecture, we highlight the key tradeoffs and research challenges at each layer.

\begin{figure}[ht]
    \centering
    \includegraphics[width=.7\linewidth]{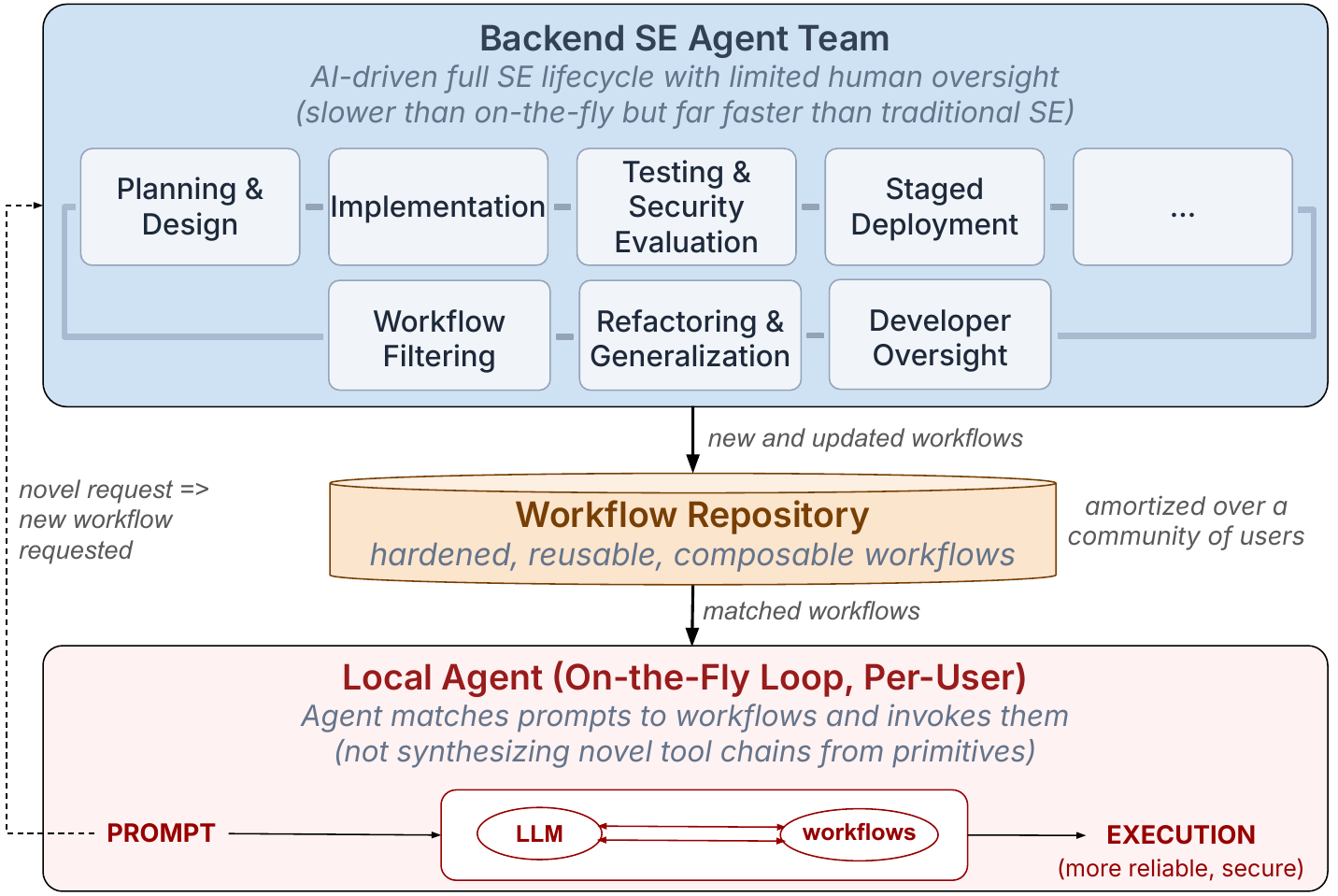}
    \caption{\small The AI Workflow Store architecture.}
    \label{fig:architecture}
\end{figure}
Figure~\ref{fig:architecture} shows the three layers of our envisioned architecture. At the top, a {\em backend SE agent team} runs the
SE lifecycle: when no workflow exists that can serve the user's request, the
agent team engineers a new one through a full SE lifecycle, potentially with
different agents specialized for different steps. The team aims not just to
solve the immediate request, but rather to create {\em general} workflows so that future requests---and future workflow production
itself---become faster, cheaper, and equally robust. 
In the middle, a {\em workflow repository} stores these general, hardened workflows and makes them available across users and agents. Accordingly, beyond
traditional SE processes, the SE agent team also performs repository maintenance
tasks such as filtering, deduplication, refactoring, and generalization. 
At the
bottom, a user's {\em local agent} resolves requests on the fly, primarily by
matching them with workflows in the repository and invoking them, rather than
always synthesizing novel and potentially complex tool chains from primitive APIs.

Consider the example from \S\ref{sec:example}. When the request to ``book the Airbnb
hotel that Bob recommended in his recent email'' reaches the local agent,
let us assume no matching workflow exists. The local agent thus forwards the request 
to the backend, where the SE agent team launches the SE lifecycle to produce
several variants to evaluate and compare in
production.\footnote[1]{\S\ref{sec:vision:local-agent} describes how the local
agent can balance flexibility and latency requirements with robustness by
supporting both synchronous and asynchronous workflow production modes.} To maximize
reuse, the development task abstracts out the user's own specific request to a
more generic form, such as \emph{building a secure connector between a
recommendation in a
generic untrusted communication channel and a high-stakes action on a booking
platform}. Once the workflow is produced and vetted, it enters the repository.
From that point on, the local agent serves any related requests, from the same
user or from others, by invoking the stored workflow rather than re-engineering
it. For example, the same workflow would serve a later request like ``book the
hotel that Alice recommended over WhatsApp.''

The top layer running the SE lifecycle is expensive but runs once per novel
task; on-the-fly matching and invocation of workflows is cheap and runs once per
user request. If our hypothesis is correct, the overhead from proper workflow
engineering---relative to creating an on-the-fly prototype---amortizes
across the user community each workflow eventually serves, transforming a
prohibitive per-request cost into a durable shared asset. In the long run,
producing reusable workflows may yield not only more secure and reliable agentic
AI, but also {\em faster and cheaper} execution---an added bonus of our vision.

The remainder of this section surfaces some of the tradeoffs and challenges we
anticipate for the design of each component, beginning with the design of the
workflow abstraction itself.

\input{03.1-workflow}
\input{03.2-backend-ai-agent-team}
\input{03.3-workflow-repository}
\input{03.4-local-agent}

%% file: 03.1-workflow.tex
\subsection{What is a workflow?}  %
\label{sec:workflow}

Each \emph{workflow} acts as an abstract intermediate representation for a set of
\emph{workflow instantiations} (\ie agent executions). Each novel request
triggers the backend to generate a corresponding workflow; similar requests may
map to already-existing workflows. Given a user request, the local agent runs
a corresponding workflow instantiation. 

Workflows should provide two properties that together determine the practicality of our architecture: (1) robustness and (2) generality.
{\em Robustness} is a composite measure of the reliability (utility under benign settings, \eg LLM hallucinations) and security (utility under malicious settings, \eg prompt injections) of workflow
instantiations. 
{\em Generality} captures how many
possible user requests the workflow supports, enabling reuse and amortization of
the workflow generation and vetting cost. 
However, these properties are fundamentally at odds, posing a design
challenge of how to represent workflows in the most practical way.

On one extreme, representing a workflow as a single precise and vetted code
execution provides a robust solution but with limited generality.
On the other extreme, representing a workflow as natural language instructions
(\eg an Agent Skill~\cite{AgentSkills}) provides generality (the local
agent instantiates the workflow via LLM-generated on-the-fly executions) but
lacks robustness: the instructions provide only probabilistic
guarantees about the resulting data or control flows.

Many possible representations lie in between. For example, traditional
programs can often generalize in limited ways via parameterization or
abstract interfaces while maintaining security invariants.
Programs with limited LLM use~\cite{camel, langchain} hardcode some
deterministic program constraints to provide some data or control flow
guarantees, but increase generality by relying on LLMs to perform well-defined, limited tasks (\eg to parse an
email).
Workflow representations can also include policies~\cite{conseca, progent,
airgap} that \eg enforce specific tool call orderings or prevent particular data
or control flows, but still permit flexible execution flows.

Finally, different contexts may require workflows with different tradeoffs, even for near-identical requests (\eg a calendar agent might prioritize a flexible workflow when rescheduling a meeting with a colleague, and a more robust workflow to send a company-wide invitation). Consequently, the local agent may need to choose between different workflows for a particular request at runtime.

%% file: 03.2-backend-ai-agent-team.tex
\subsection{The SE Agent Team: Producing robust and general workflows}
\label{sec:vision:se-agent-team}

The AI Workflow Store mandates that the backend SE agent team reintroduce SE
rigor to the personal-agent world, by executing the full SE lifecycle for each new
workflow and producing durable, shared artifacts meant to be invoked by many
users across many sessions. %
This mandate is intentionally ambitious: this shift from outputting an
ephemeral trace to producing a general but hardened artifact is what makes workflow
development both promising and challenging.

In meeting this mandate, the backend team must navigate the central tradeoff 
between production cost and robustness of the produced workflows.
Producing robust workflows might entail \eg 
requirements gathering, threat modeling, exploration of multiple candidate
designs, implementation, adversarial testing, staged deployment, and continuous
refactoring and generalization as related workflows accumulate. 
Ideally, this expensive work would pay off as local agents, across many users,
avoid repeating it and benefit from the robustness of the produced workflow.
In reality, if pushed too far, production costs can exceed any realistic amortization
benefit; and if pushed too little, the resulting workflows may be no more robust
than on-the-fly synthesis. 

Where this balance ends up depends on how cheaply each SE step can be automated.
Coding agent teams present a natural starting point: they already demonstrate
the ability to autonomously apply some disciplined SE processes when given time
and iteration, albeit as of yet still requiring some human
supervision~\cite{GoogleJules2026, AnthropicClaudeCode2026, cursor, codex, carlini2026building}. 
However, two steps---design exploration and adversarial testing---pose
particularly difficult automation challenges.
Creative design exploration may require human intervention or access to
established design patterns. How, for example, does the agent team discover the
``Add to Calendar''-style abstraction for hotel recommendations, a design that is very different
from what an on-the-fly agent would likely attempt? Adversarial testing raises a
similar concern: automated security evaluation is only as strong as the attacks
the system can generate or retrieve. 
Our vision thus depends on the development of datasets of rich patterns for both
constructive design and offensive testing. Without them, backend agents risk
producing workflows robust only to the narrow cases they can consider---precisely the
failure mode we seek to escape. 

Unlike much of the personal-agents literature, which treats human involvement as
something to minimize or avoid, our architecture explicitly supports it. When
carefully scoped and applied to high-leverage points, such involvement can
meaningfully improve both quality and
safety~\cite{DBLP:journals/corr/abs-2502-02649}. For example, expert developers
can review, intervene in, and approve workflows at production time, providing a
final layer of accountability for what enters the workflow repository and a
defense against malicious or low-quality outputs. Involving end-users to clarify
goals, gather examples, and co-develop a precise specification for a novel
request mirrors traditional requirements gathering and avoids guesswork. This
process can aggregate input across users requesting similar workflows and
trigger production only once sufficient detail is available. Finally, both
expert developers and volunteer users can play a role in maintaining and
improving the workflow repository over time.

%% file: 03.3-workflow-repository.tex
\subsection{The Workflow Repository: Maintaining and evolving workflows}
\label{sec:vision:workflow-repository}

The workflow repository continuously collects and provides workflows to serve a community of users.
Unlike on-the-fly agent executions, which are
unstructured and
isolated, workflows are structured and versioned artifacts amenable to more
effective engineering processes at scale. 
For example, each workflow reuse produces additional real-world traces and
feedback from participating end-users that help the backend SE team surface
edge cases and harden them further, improving robustness.
The SE team performs continual workflow optimizations offline on the
collective workflows of all users (\eg maintenance, deduplication,
generalization, composition, and improvement of workflows).

However, these benefits come with the added cost of more backend SE
team engineering and leads to \emph{tradeoffs between all of robustness,
generality, and cost.} 
Maintaining workflows over long horizons might require the backend team to
refactor and handle workflows with deprecated or evolving APIs;
and the SE team must vet the robustness of any generalized or composed
workflows. 
Generalization is critical to practically amortizing these SE costs with reuse
(\S\ref{sec:vision:se-agent-team}), although generality may be at odds with robustness (\S\ref{sec:workflow}).

A shared repository design must also handle the threat of attacks mirroring those on
code repositories, shared caches and package-manager ecosystems: (1) {\em
malicious injection}---where the adversary tricks the backend into producing
backdoored workflows; (2) {\em spurious requests}---resource exhaustion attacks
targeting the SE agent team; (3) {\em discovery hijacking}---where illegitimate
workflows mask trusted ones to redirect matching traffic; and (4) {\em
side-channel attacks}---where lookup latency or leaks from users' initial
requests reveal user activity across the shared store. While these reflect
classic threats, their agentic instantiations may require domain-specific
solutions.

%% file: 03.4-local-agent.tex
\subsection{The Local Agent: Matching and executing workflows---a flexibility/robustness choice}
\label{sec:vision:local-agent}

Users interact with the AI Workflow Store via a \emph{local
agent} that (1) matches requests to repository workflows or escalates to the SE agent team, and (2) instantiates and executes the selected workflow. As discussed in \S\ref{sec:workflow}, workflow representation shapes generality and robustness. The local agent's matching, escalation, and execution logic further determines both.

\input{03.4-local-agent-execution-models}

The local agent can structure a workflow's execution in various ways, raising a key design question.
Listings~\ref{lst:workflow-rigid}, ~\ref{lst:workflow-flexible}
and~\ref{lst:workflow-chained-async} illustrate three variants, contrasted with
the on-the-fly loop in Listing~\ref{lst:onthefly}. They differ along two axes
that together navigate a flexibility-robustness spectrum: single vs. chained
workflows, and synchronous vs. asynchronous workflow escalation.

At the rigid extreme (Listing~\ref{lst:workflow-rigid}), each user request maps
to a single, pre-built workflow, instantiated from the prompt and executed
without chaining or cross-invocation state. If no match exists, the agent waits
or refuses the request. This mirrors systems like Conseca~\cite{conseca} and
CaMeL~\cite{camel}, but workflows go further: they offer more robustness (produced via
SE pipelines) and potentially more generality (via intentional preparation of
generic, possibly multi-stage plans, which Conseca and CaMeL do not support).
Compared to the vanilla on-the-fly loop, however, they reduce flexibility: the
local agent cannot serve unmatched requests immediately.

A more flexible design (Listing~\ref{lst:workflow-flexible}) allows chaining
workflows in a loop, with each step extending the context. This improves
flexibility in solving more complex tasks, but weakens robustness: while the SE
agent team vets individual workflows, they may not vet workflow chains. This
begins to resemble tool-based on-the-fly agents, where integration bugs arise
despite potentially well-tested tools. Still, workflows should offer an
advantage in robustness here: they can be thought of as encapsulating
higher-level, continuously-refined chains of tools, raising the abstraction
level of on-the-fly, tool-based execution.

The local agent can push flexibility further by allowing asynchronous workflow
production (Listing~\ref{lst:workflow-chained-async}). When no match exists, the
agent registers the request for backend synthesis and immediately proceeds with
an on-the-fly execution. This sacrifices robustness for the current
request---degrading it to the level of vanilla on-the-fly agents---but provides 
robustness for future executions once the workflow is produced.

This design raises several challenges. Correct matching is essential---workflows
must only be invoked within their tested envelope---and might be harder than
tool selection~\cite{tooluse_anthropic,tooluse_openai} due to the scale and
potential overlap of workflows. The agent must also choose an appropriate point
on the flexibility-robustness spectrum per request; for instance, it might
choose between correctly-matched workflows that offer different properties, or
decide how best to execute the workflow (chained, synchronous, etc.).

The AI Workflow Store architecture helps with these choices: it narrows choices to
higher-level units,
enables deduplication and monitoring, matches on trusted input before exposing
workflows to untrusted context, and can leverage workflow annotations regarding
the testing envelope or the level of vetting to determine if an invocation may
be safe. More generally, the biggest advantage we anticipate from our approach
is that it creates a navigable design space with benefits afforded by proper
engineering that other solutions cannot reach within the rushed, on-the-fly
tool orchestration model. After reviewing additional such solutions, we return
to this design space and the AI Workflow Store's place in it vis-a-vis other
works in \S\ref{sec:broader-perspective}.

%% file: 03.4-local-agent-execution-models.tex
\begin{figure*}[htb]
\centering \footnotesize
\vspace{-.5cm}
\begin{minipage}[b]{.48\textwidth}
\begin{lstlisting}[language=Python, frame=none, basicstyle=\footnotesize\ttfamily, columns=flexible, breaklines=true, numbers=none, caption={\small Vanilla on-the-fly agent}, label={lst:onthefly}]
      context = getPromptAndContext() 
      while llm.continue(context):
        # Get next tool to call on-the-fly
        tool = llm.getNextTool(context)
        out = tool.exec(context)
        context.add(out)
      returnToUser(getOutput(context))
\end{lstlisting}
\end{minipage}\hfill
\begin{minipage}[b]{.48\textwidth}
\begin{lstlisting}[
    language=Python, frame=none, basicstyle=\footnotesize\ttfamily, columns=flexible, breaklines=true, numbers=none, 
    caption={\small Single sync. workflow (rigid extreme)}, label={lst:workflow-rigid}]
    context = getPromptAndContext()
    # Retrieve and execute a single workflow
    workflow = store.getWorkflow(context)
    if not workflow:
      workflow = store.engineerNewWF(context)
    out = workflow.exec(context)
    returnToUser(out)
\end{lstlisting}
\end{minipage}
\hfill

\begin{minipage}[b]{.48\textwidth}
\begin{lstlisting}[
    language=Python, frame=none, basicstyle=\footnotesize\ttfamily, columns=flexible, breaklines=true, numbers=none, 
    caption={\small Chained sync. workflows (more flexible)}, label={lst:workflow-flexible}]
    context = getPromptAndContext()
    while llm.continue(context):
      # Retrieve next workflow based on 
      # growing context
      workflow = store.getWorkflow(context)
      if not workflow:
        workflow = store.engineerNewWF(context)
      out = workflow.exec(context)
      context.add(out)
    returnToUser(getOutput(context))
\end{lstlisting}
\end{minipage}
\begin{minipage}[b]{.48\textwidth}
\begin{lstlisting}[
    language=Python, frame=none, basicstyle=\footnotesize\ttfamily, columns=flexible, breaklines=true, numbers=none, 
  caption={\small Async. workflows (flexible extreme)}, label={lst:workflow-chained-async}]
  context = getPromptAndContext()
  while llm.continue(context):
    workflow = store.getWorkflow(context)
    if not workflow:
      store.asyncEngineerNewWF(context)
      # Vanilla tool-calling loop, no wait
      out = execOnTheFly(context) 
    else:
      out = workflow.exec(context)
    context.add(out)
  returnToUser(getOutput(context))
\end{lstlisting}
\end{minipage}
\vspace{-10pt}
\end{figure*}

%% file: 04-related.tex
\section{Related Work}
\label{sec:related}

The AI Workflow Store restricts possible agent behaviors by engineering hardened workflows, improving reliability and security. 
Orthogonal approaches instead add deterministic constraints by
enforcing static policies~\cite{buhler2026agentbound,
sharma2026ac4a, pcas, provos2026ironcurtain, nanoclaw, cloudflare} or policies
generated ``on-the-fly''~\cite{conseca, camel, ace, airgap, isolategpt} while
isolated from untrusted context~\cite{willisonisolation}. 
Static policies often end up too coarse-grained or rigid to adapt to the many
contexts experienced by personal agents~\cite{conseca}. On-the-fly ones
generated by isolated LLMs can be fine-grained, but assume that untrusted data
serves as the only way to introduce vulnerabilities. In practice, however, this
threat model is unrealistic: even if the LLM is isolated to trusted
context~\cite{willisonisolation}, the agents can introduce security
vulnerabilities even in the absence of adversarial data~\cite{perry2023users}
and users might unwittingly ask the agent to do something unsafe (\eg skip
authentication checks).
The AI Workflow Store reintroduces traditional SE processes like requirement design
and red-teaming precisely to address such issues (see discussion in
\S\ref{sec:example:defense}).

Some policy systems~\cite{miniscope, stanley2026ai, prudentia} allow users to interact
on-the-fly to specify agent permissions. Workflows can be engineered to involve
user interaction, but we predict many such user preferences and requirements can be determined inside the SE lifecycle that produces the workflow.
Others add information flow control policies to catch mistaken on-the-fly
dataflows~\cite{camel, fsecure, fides, prudentia, stanley2026ai}; because workflows are
engineered, static analysis can determine if the workflow adheres to these
policies, and dynamic taint tracking could be added at runtime as an additional
layer of assurance.
Finally, some approaches add LLM-based ``guardrails'' for robustness without isolation from untrusted context~\cite{llmasajudge, llamaguard, llamafirewall, zverev2025aside, chen2025secalign, progent} (\eg to detect prompt injections or check for action safety or security). 
These approaches cannot provide any guarantees in the face of prompt injections
due to LLMs' inherently probabilistic nature, and are best complemented
with deterministic constraints.

Closest to our proposal are an emerging class of {\em reusable components}, such as {\em skills}.
Examples include the old Alexa Skills~\cite{AlexaSkills} and Google Home Services~\cite{GoogleHomeServices}, which packaged pre-engineered, voice-invoked functionality as reusable modules;
and newer agent skills~\cite{AgentSkills}, which encode task-specific expertise (typically prompts) to be incorporated by AI agents as additional context.
% with some systems such as OpenClaw generating and saving them automatically~\cite{openclaw-skills}.
Such components can be created manually~\cite{SkillGuide} or with help from AI~\cite{SkillCreator}, and are often shared in public repositories~\cite{SkillsMP,AnthropicSkills,AwesomeOpenClawSkills,ClawHub}.
Recent tools also help review skills for security risks~\cite{nvidia}, or compile skills into structured programs with LLM invocation~\cite{mellea}; this compilation could act as the first step in producing a possible representation of workflows as structured programs.
Invoking such prebuilt components can improve reliability, reduce errors, and conserve resources for agents and other runtime systems.
Stoica et al.~\cite[\S8.4]{stoica2024specifications} also identify reusability as a core property that LLM systems should strive for, while Skyvern~\cite{Skyvern} automates browser ``workflow'' creation from demonstrations to provide a more reliable alternative to hallucination-prone, on-the-fly browser agents. Finally, enterprise and coding agents~\cite{GoogleJules2026, AnthropicClaudeCode2026,cursor,codex} typically already incorporate the types of rigorous SE and preproduction processes we argue for in this paper, though they remain primarily human-driven.

The AI Workflow Store vision pushes these ideas further and to the scope of personal agents specifically, proposing a concrete architecture and automatic ecosystem in which workflows are robustly engineered, discovered, reused, and invoked {\em as the norm for personal-agent execution}.

%% file: 05-broader-perspective.tex
\section{Broader Perspective}
\label{sec:broader-perspective}

\begin{figure}[h]
    \centering
    \includegraphics[width=0.8\linewidth]{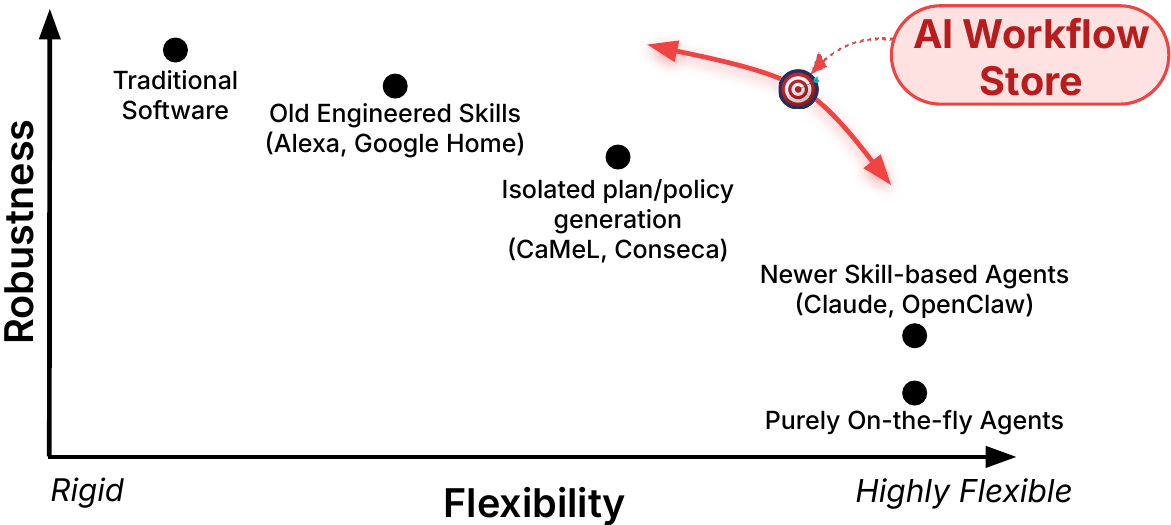}
    
    \caption{The flexibility/robustness spectrum, with existing solutions and our proposed vision.}
    \label{fig:flexibility-robustness-spectrum}
\end{figure}

Figure~\ref{fig:flexibility-robustness-spectrum} positions our vision within the spectrum defined by the tension between {\em flexibility} (ability to respond to any user need with the right functionality) and {\em robustness} (reliability and security of that functionality). Traditional software sits at one extreme: highly robust through careful engineering, but expensive to produce and limited in scope and flexibility. Purely on-the-fly agents sit at the other extreme: highly flexible but fragile. Prior approaches occupy intermediate points: engineered skills (e.g., Alexa, Google Home) offer robustness but limited flexibility due to their developer-driven production, while newer skill-based agents like OpenClaw offer high flexibility but mostly follow the on-the-fly paradigm with limited robustness. 
Approaches for isolated (and trusted) plan or policy generation~\cite{camel,
conseca} add robustness (better security against prompt injection) but lack
explicit engineering and the flexibility to handle multi-stage executions; they
likely exhibit robustness in-between engineered skills and the new generation of skills. 
The AI Workflow Store pushes toward the ideal top-right corner, increasing robustness through proper engineering while recovering flexibility via automated production.

Stepping back, we give our perspective on why navigating this tension requires moving beyond on-the-fly execution toward engineered workflows. Consider what happens as we move from traditional software to agentic systems: this essentially renegotiates the boundaries between developer control, functional flexibility, and robustness. 
Increasing application ``agenticness''---where the developer offloads more
application development to AI---expands flexibility but reduces developer
control, creating a gap between developer intent and application specification.
While LLMs excel at general-purpose reasoning, their only true specification is
to transform input tokens into output tokens; as of yet, we have no guarantee
they will correctly realize developer intent, even when carefully prompted.

In traditional software, alignment between developer intent and implementation
is achieved through disciplined software engineering---not by
brilliance, but by process. We argue that robustness in agentic systems can be
similarly achieved by integrating SE processes---design, testing, adversarial
evaluation, and iteration---into the agentic loop. The AI Workflow Store
operationalizes this shift by replacing per-request improvisation with hardened,
reusable workflows and amortizing the engineering effort across users and requests.
As agents increasingly enter high-stakes settings, on-the-fly execution alone is unlikely to
provide the level of robustness they require---%
a gap we aim to close by treating robustness not as an emergent property, but as an \emph{engineered} property.

%% file: 07-acks.tex
\section*{Acknowledgements}

This paper benefited from the conversations, feedback, and work of many. We are
particularly grateful to: Adria Gascon, for highlighting the resource asymmetry
between attackers and on-the-fly agents, where a lack of defensive iteration
creates an inherent security disadvantage; Marco Gruteser, for his insights into
the design spectrum between early, human-developed ``skills'' (\eg Alexa, Google
Home Services) and contemporary agentic processes, and where our vision lands on
this spectrum; Greg Ganger, Kim Keeton, and Hank Levy for help with ideation from
the start and feedback throughout; Mihai Christodorescu and David Culler for their
feedback that guided this paper's development; Alex Krentsel for his insights on
OpenClaw architecture and skills; and Jorge Ortiz, Simha Sethumadhavan, and
members of the Columbia Software Systems group for inspiring conversations during
a reading group on agentic AI security and privacy.